\documentclass{article}
\usepackage{frascatiphys}
\usepackage{graphicx}

\begin{document}
\title{ 
CP VIOLATION IN CHARM
}
\author{
Kevin Stenson        \\
\textit{Dept of Physics \& Astronomy, VU Station B 351807,}\\
\textit{Vanderbilt University, Nashville, TN 37235 USA} \\
}
\maketitle
\baselineskip=11.6pt
\begin{abstract}
Recent results on searches for CP and CPT violation in the charm sector are presented.
These results include limits on direct CP violation in several channels from the FOCUS and
CLEO experiments.  The first reported search for CPT violation in charm, 
a preliminary result by the FOCUS collaboration, is also presented.
\end{abstract}
\baselineskip=14pt
\section{Charm CP Violation Introduction}
CP violation is generally divided into three types: CP violation in mixing (indirect),
CP violation in decay (direct), and CP violation in the interference between 
decay and mixing (indirect or direct).  
In all cases, CP violation occurs when the decay rate of a particle differs
from that of its CP conjugate.  This requires contributions from two different CP 
violating terms with different phases.  In addition, two CP conserving terms must
also have different phases.  The CP conserving phase shift is usually generated by
QCD final state interactions.  In the Standard Model (SM), two CP violating terms often come
from tree level and penguin decays.  
Extensions to the Standard Model can introduce other CP violating terms which
can interfere with the SM weak decays to generate CP violation.

In charm, mixing is very suppressed so at current
experimental sensitivities, CP violation searches are generally searching for direct CP 
violation.  One measures the CP violation rate by looking at the asymmetry:
\begin{equation}
A_{CP} \:\equiv\: \frac{\Gamma(D \rightarrow f) \:-\: \Gamma(\overline{D} \rightarrow \overline{f})}
 {\Gamma(D \rightarrow f) \:+\: \Gamma(\overline{D} \rightarrow \overline{f})}
\end{equation}
In the fixed-target experiments E791 and FOCUS, the production mechanism gives rise to different 
numbers of produced particles and antiparticles.  Therefore, these experiments normalize to another 
(copious) decay mode which is unlikely to exhibit CP violation.


\section{Overview of Experiments}
The most precise charm CP violation results come from the Fermilab fixed-target experiments E791 and FOCUS 
and the $e^+e^-$ central detector, CLEO.
\subsection{E791 and FOCUS experiments}
E791 (FOCUS) took data at Fermilab during the fixed-target running of 1991--2 (1996--7).  These experiments,
like all modern fixed-target charm experiments are quite similar.  Both sport silicon strip detectors in
the vertex region to separate the charm production and decay vertices, a key requirement in separating
signal from background.  Following the silicon detectors
are wire chambers and magnets which track and momentum analyze the decay products.  Particle identification
of charged hadrons is accomplished by the use of 2 (E791) or 3 (FOCUS) multi-cell 
threshold \v{C}erenkov counters.  Electromagnetic calorimeters identify electrons and photons while 
scintillation counters downstream of absorbing steel walls are used to identify muons.  Both experiments
used a hadron calorimeter to trigger on interesting events with high efficiency.  The targets in both
experiments were segmented to allow charm decays in air.  E791 used a 500~GeV/$c$ $\pi^-$ beam 
while FOCUS used a photon beam with an average energy of 180 GeV (for events with a reconstructed charm
particle).  The average charm momentum was around 60~GeV/$c$ for both experiments.
From a collection of 20 billion (6 billion) triggered events, E791 (FOCUS) fully reconstructed more than
200,000 (1,000,000) charm particles.

\subsection{CLEO experiment}
The CLEO experiment utilizes the CESR storage ring at Cornell which is a symmetric $e^+e^-$ collider.  
The CLEO results presented here come from data taken at and near the $\Upsilon(4S)$, mostly from 
CLEO II.V (1996--9).  Both CLEO II\,\cite{CLEO2} and CLEO II.V\,\cite{CLEO25} detectors use wire 
chambers for particle tracking and an
excellent electromagnetic CsI calorimeter providing good reconstruction of photons, electron, and $\pi^0$'s.  
These detectors are inside a 1.5~T axial magnetic field and surrounded by muon chambers.  In CLEO II.V a 
silicon strip system near the beam was also present.  The data presented here utilize 4.7--13.7~pb$^{-1}$ of
luminosity. Charm particles produced at CLEO generally have a momentum of a few GeV/$c$.

%
\section{Direct CP Violation Results}
\subsection{Two-body decays}

\begin{figure}[htbp]
\includegraphics[width=4.7in,height=3.0in]{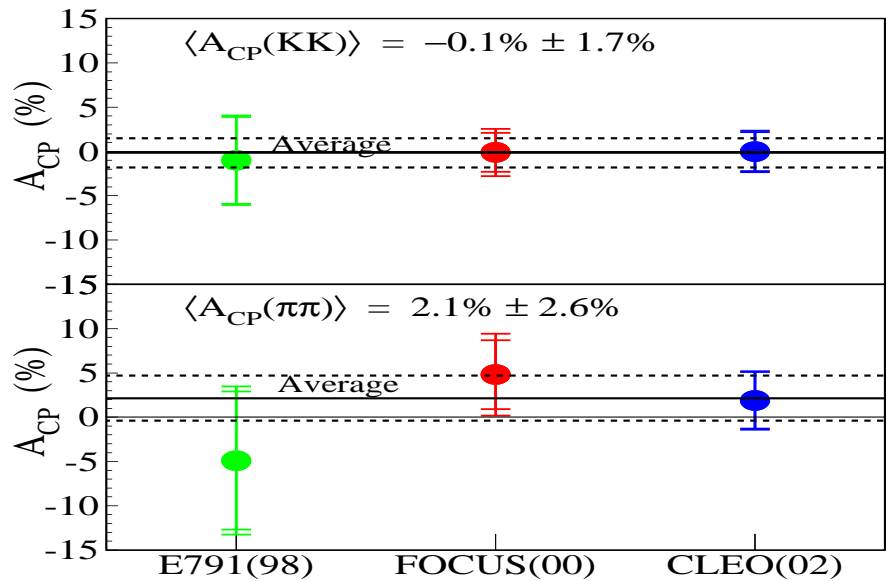}
 \caption{\it Measurements of the CP asymmetry of $D^0\!\rightarrow\! K^+K^-,\; \pi^+\pi^-$
decays.
    \label{fig:cpviol_d0} }
\end{figure}

E791\,\cite{e791_cpviol_d0}, FOCUS\,\cite{focus_cpviol_d0}, and CLEO\,\cite{cleo_cpviol_d0_chg} have all 
looked for CP violating behavior in the Cabibbo suppressed decays $D^0\!\rightarrow\! K^+K^-$ and 
$D^0\!\rightarrow\! \pi^+\pi^-$.  These measurements, shown in Fig.~\ref{fig:cpviol_d0} and tabulated
in Table~\ref{tab:cpviol_d0}
are approaching the 1\% level where non-Standard Model effects might show up.  

\begin{table}[htbp]
\centering
\caption{\it Measurements of the CP asymmetry from $D^0\!\rightarrow\! K^+K^-,\; \pi^+\pi^-$
decays.}
\vskip 0.1 in
\begin{tabular}{|l|c|c|} \hline
Expt & $A_{CP}(KK)$ (\%) & $A_{CP}(\pi\pi)$ (\%) \\ \hline
E791(98)\,\cite{e791_cpviol_d0}   & $-1.0 \pm 4.9 \pm 1.2$ & $-4.9 \pm 7.8 \pm 3.0$ \\
FOCUS(00)\,\cite{focus_cpviol_d0} & $-0.1 \pm 2.2 \pm 1.5$ & $\;\;\;4.8 \pm 3.9 \pm 2.5$ \\
CLEO(02)\,\cite{cleo_cpviol_d0_chg} & $\:\;\;0.0 \pm 2.2 \pm 0.8$ & $\;\;\;1.9 \pm 3.2 \pm 0.8$ \\ \hline
\end{tabular}
\label{tab:cpviol_d0}
\end{table}

FOCUS has recently published\,\cite{focus_cpviol_ks} results using the two-body decay modes $D^+\!\rightarrow\! K_S^0 \pi^+$, where 
Cabibbo favored and doubly Cabibbo suppressed amplitudes can interfere and 
$D^+ \!\rightarrow\! K_S^0 K^+$ which is a singly Cabibbo suppressed decay where interference between the tree
and penguin diagrams can occur.  These decay modes should also display a CP violation component due to CP violation
in $K^0$ decays.  As seen in Table~\ref{tab:cpviol_dp_ks}, no evidence of CP violation was found which is consistent 
with the Standard Model for this level of sensitivity.

\begin{table}[htbp]
\centering
\caption{\it Measurements of the CP asymmetry from $D^+ \!\rightarrow\! K_S^0 K^+,\: K_S^0 \pi^+$
decays.}
\vskip 0.1 in
\begin{tabular}{|l|c|} \hline
CP Asymmetry & {FOCUS} \\ \hline
$A_{CP}(K_{\!S}^0 \pi^+)$ w.r.t. $K^- \pi^+ \pi^+\!\!$ & {$(-1.6 \!\pm\! 1.5 \!\pm\!0.9)$\%$\!\!$} \\
$A_{CP}(K_{\!S}^0 K^+)$ w.r.t. $K^- \pi^+ \pi^+\!\!$ & {$(6.9 \!\pm\! 6.0 \!\pm\!1.8)$\%$\!\!$} \\
$A_{CP}(K_{\!S}^0 K^+)$ w.r.t. $K_{\!S}^0 \pi^+$ & {$(7.1 \!\pm\! 6.1 \!\pm\!1.4)$\%$\!\!$} \\ \hline
\end{tabular}
\label{tab:cpviol_dp_ks}
\end{table}

\subsection{Three-body decays}

Searching for direct CP violation in three-body decays is significantly more complicated than two-body decays.
One can look for CP violation by integrating over phase space, by looking at quasi two-body decays by cutting on
resonances, or by using a full Dalitz plot analysis of the charm and anticharm particle to look for discrepancies.
FOCUS\,\cite{focus_cpviol_d0} and E791\,\cite{e791_cpviol_dp} have both reported results for 
$D^+ \!\rightarrow\! K^-K^+\pi^+$.  FOCUS reported $A_{CP} = (0.6 \pm 1.1 \pm 0.5)\%$ and plans to perform a CP violation Dalitz
plot analysis in the future.  E791 also reported results for the sub-resonances: $A_{CP}(\phi\pi^+) = (-2.8 \pm 3.6)\%$ and
$A_{CP}(K^*K^+) = (-1.0 \pm 5.0)\%$.

As a byproduct of their Dalitz plot analysis of $D^0\!\rightarrow\! K^-\pi^+\pi^0$\,\cite{cleo_kpipi0}, CLEO measured
\begin{equation}
A_{CP} \; \equiv \; \int \frac{\left| M_{D^0}\right|^2 - \left| M_{\overline{D}\,\!^0}\right|^2}
{\left| M_{D^0}\right|^2 - \left| M_{\overline{D}\,\!^0}\right|^2} \,d D\!P \;=\; (-3.1 \pm 8.6)\%,
\end{equation}
again consistent with zero.

\section{CPT Violation Search}

It is common knowledge that point particle Lorentz invariant field theories require CPT invariance\,\cite{sachs}.
However, some Standard Model extensions need not be Lorentz-invariant\,\cite{cpt_physics}.  In fact, it might be possible
to find evidence for strings which dominate at the Plank scale using data which exists today. Limits on
CPT violation have been set using neutral $K$ and $B$ mesons (mixing interferometry).  It is possible,
however, for these effects to manifest at different levels in different flavors so a check in the charm
system is also important.

\subsection{CPT Violation Formalism}
This analysis mostly follows the notation of Ref.~\cite{cpt_formalism}.  First, the standard effective 
Hamiltonian is rewritten:
\begin{equation}
\Lambda \;=\; M - \frac{1}{2}i\Gamma \;\;\;\;\;\;\; \Longrightarrow \;\;\;\;\;\;\; \Lambda \;=\; \frac{1}{2} \,\Delta \lambda \left( 
\begin{array}{ccc}
U \!+ {\xi} & \; & V\, W^{-1} \\
V\, W & \; & U \!- {\xi} \\
\end{array}
\right)
\end{equation}
where $U$, $V$, $W$, and $\xi$ are complex and $\Delta \lambda = \Delta M - i\, \Delta \Gamma / 2$.
The parameter $\xi$ is the CPT violating term.
The time-dependent right-sign $D^0 \rightarrow f$ decay probability is given by:
\begin{eqnarray}
P_f(t) & = & \frac{1}{2}|F|^2 e^{-\Gamma t}
\left[ \left(1 \!+\! |{\xi}|^2\right) \cosh \Delta \Gamma + 
\left(1 \!-\! |{\xi}|^2\right) \cos \Delta M \right. \nonumber \\
       &  & \mbox{} - \left. 2 \Re({\xi}) \sinh \Delta \Gamma + 2 \Im({\xi}) \sin \Delta M \right].
\end{eqnarray}
The time-dependent $\overline{D}\!\,^0 \rightarrow \overline{f}$ decay probability $\overline{P}_{\overline{f}}(t)$ is 
simply $P_f(t)$ with ${\xi} \rightarrow -{\xi}$ and $F \rightarrow \overline{F}$.
From this, one can form an asymmetry for right-sign decays as:
\begin{equation}
A_{C\!P\!\!\;T}(t) \;=\; \frac{\overline{P}_{\overline{f}}(t) - P_f(t)}{\overline{P}_{\overline{f}}(t) + P_f(t)}
\;=\; \frac{2 \Re({\xi}) \sinh \Delta \!\Gamma t - 2 \Im({\xi}) \sin \Delta\! M t}
{\left(1 \!+\! |{\xi}|^2\right) \cosh \Delta \!\Gamma t + 
\left(1 \!-\! |{\xi}|^2\right) \cos \Delta\! M t}.
\end{equation}
By Taylor expanding $\sin, \sinh, \cos, \cosh$ to 1st order and switching to the standard mixing variables
$x \equiv \Delta M \!/\Gamma$, $y \equiv \Delta \Gamma \!/ 2\Gamma$, one finds:
\begin{equation}
A_{C\!P\!\!\;T}(t) \;\approx\; \left[\Re({\xi})\, y - \Im({\xi})\, x)\right] \Gamma t
\end{equation} 
Given the measured limits on mixing and the lifetime range probed by the {FOCUS} experiment, 
this approximation is sufficiently accurate.
Experimentally:
\begin{equation}
A_{C\!P\!\!\;T}(t') \;=\; \frac{N_{\overline{D}\!\,^0}(t') - N_{D^0}(t')}{N_{\overline{D}\!\,^0}(t') + N_{D^0}(t')}.
\end{equation}
Therefore, measuring the slope of the lifetime ratio distribution
immediately returns $\left[\Re({\xi})\, y - \Im({\xi})\, x\right]$.

\subsection{Preliminary FOCUS CPT Violation Results}
\begin{figure}[htbp]
\includegraphics[width=4.7in,height=2.75in]{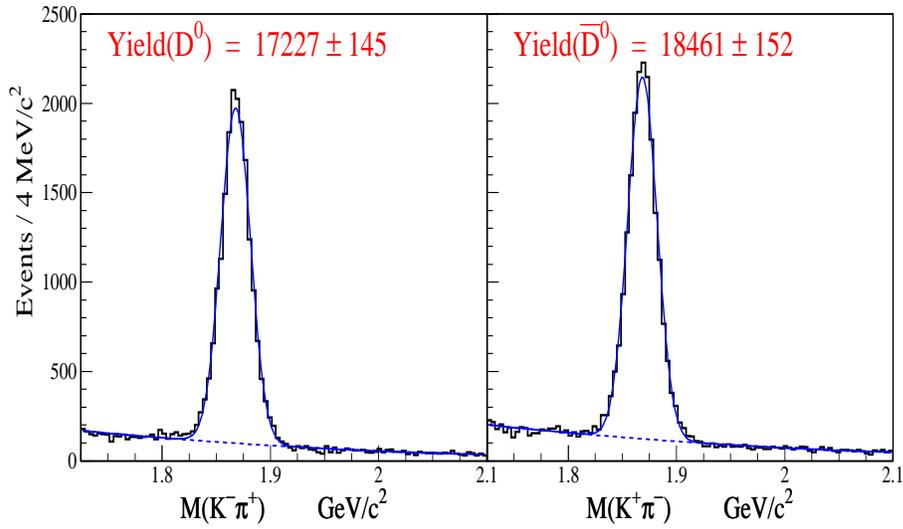}
 \caption{\it Preliminary FOCUS mass plots of $D^0 \!\rightarrow\! K^- \pi^+$ for events which have
a $D^*$ tag.
    \label{fig:cpt_mass} }
\end{figure}

Figure~\ref{fig:cpt_mass} shows the invariant mass distribution for right-sign $D^0 \!\rightarrow\! K^-\pi^+$ decays.
These decays have been tagged using the charge of the soft pion from $D^{*+} \!\rightarrow\! D^0 \pi^+_s$ decays.

\begin{figure}[htbp]
\includegraphics[width=4.7in,height=3.0in]{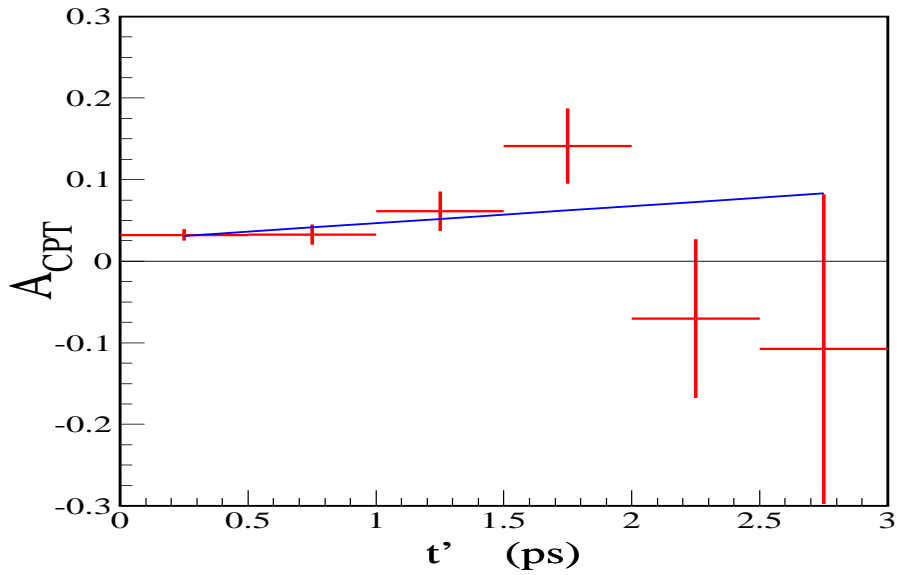}
 \caption{\it Preliminary FOCUS fit to the ratio of $\overline{D}\!\,^0$ to $D^0$ as a function of reduced
proper time, $t'$.
    \label{fig:cpt_fit} }
\end{figure}

In Fig.~\ref{fig:cpt_fit}, the ratio of $\overline{D}\!\,^0$ to $D^0$ as a function of reduced
proper time, $t'$, is plotted.  The reduced proper time is defined by $t' \equiv (\ell - N \sigma_\ell)/(\beta \gamma c)$ 
where $\ell$ is the distance between the production and decay vertex, $\sigma_\ell$ is the calculated resolution on
$\ell$, and $N$ is the minimum detachment cut applied.  This has the effect of starting the clock at the moment at which
the particle could first be reconstructed by the experiment (with the given detachment cut) and thus greatly reducing the
amount of correction needed due to acceptance.  The fit to 
Fig.~\ref{fig:cpt_fit} is the basis of the preliminary FOCUS result:
\begin{equation}
\Re(\xi)\,y \:-\: \Im(\xi)\,x \;\:=\:\; 0.0083 \,\pm\, 0.0065 \,\pm\, 0.0041
\end{equation}
The actual limit on the CPT violating parameter depends on mixing parameters; for example 
if $x=0$ and $y=1\%$ then $\Re(\xi) = 0.83 \pm 0.65 \pm 0.41$.  The systematic errors were determined
by exploring the effect of different absorption lengths in the Monte Carlo simulation, different selection
criteria, and different sideband selections for the background subtraction.

\section{Conclusion}

Considerable progress is currently being made by FOCUS and CLEO in the search for direct CP violation
in the charm system.  Current limits are approaching the 1\% level at which non-Standard Model effects might
be seen.  In the near future, BaBar and Belle should be able to probe even further in this exciting area.
Following this, BTeV will also have an opportunity to search for CP violation in the charm sector with a sample
of more than 1 billion reconstructed charm decays.  In addition, the search for CPT violation is also being
extended into the charm sector.  Both types of searches have the potential to uncover exciting new physics.

\end{document}